%% file: main.tex
\documentclass{article}
\usepackage[T1]{fontenc}
\usepackage{c_style,times}

\input{Text/math_commands.tex}

\input{Text/definition}

\usepackage{hyperref}
\hypersetup{
  colorlinks=true,
  linkcolor=blue,
  citecolor=blue,
  urlcolor=blue
}
\usepackage{url}
\usepackage{cleveref}

\title{\toolname{}: Should you trust the results in coding benchmarks? Agentic Coding Benchmarks Auditing via  Evaluator Evolution}

\firstpageheader{Preprint.}

\author{
\authorname{Shuangjie Yao}{1}\And
\authorname{Hao Wang}{2}\And
\authorname{Koushik Sen}{2}\And
\authorname{Simin Chen}{3}\And
\authorname{Baishakhi Ray}{1}\And
\authorname{Dawn Song}{2}
}

\affiliations{
\affiliation{1}{Columbia University}\And
\affiliation{2}{University of California, Berkeley}\And
\affiliation{3}{George Mason University}
}

\begin{document}

\maketitle

\input{Text/abst}
\input{Text/intro}
\input{Text/background}

\input{src/approach}
\input{src/evaluation}
\input{Text/conclusion}

\bibliography{misc/ref}
\bibliographystyle{c_style}

\appendix
\input{appendix/discussion}

\input{appendix/additional_experiment}

\input{appendix/case_study}

\input{appendix/prompt}

\input{appendix/limitation}

\end{document}

%% file: Text/math_commands.tex
\usepackage{amsmath,amsfonts,bm}

\def\Figref#1{Figure~\ref{#1}}

\def\eqref#1{equation~\ref{#1}}

\def\1{\bm{1}}

\DeclareMathAlphabet{\mathsfit}{\encodingdefault}{\sfdefault}{m}{sl}
\SetMathAlphabet{\mathsfit}{bold}{\encodingdefault}{\sfdefault}{bx}{n}

%% file: Text/definition.tex
\usepackage{booktabs}
\usepackage{amsfonts}
\usepackage{graphicx}
\usepackage{hyperref}
\usepackage{textcomp}
\usepackage{listings}
\usepackage{adjustbox}
\usepackage{xspace}
\usepackage{multirow}
\usepackage{multicol}
\usepackage{xcolor} 
\usepackage{amsmath}
\usepackage{tcolorbox}
\newtcolorbox{insightbox}{colback=gray!6,colframe=gray!55,boxrule=0.5pt,arc=2pt,left=5pt,right=5pt,top=3pt,bottom=3pt,before skip=6pt,after skip=6pt}
\tcbuselibrary{skins,breakable,listings}

\definecolor{promptframe}{gray}{0.32}
\definecolor{promptback}{gray}{0.965}
\lstdefinestyle{promptstyle}{
  basicstyle=\ttfamily\scriptsize,
  backgroundcolor=\color{promptback},
  numbers=none,
  frame=none,
  breaklines=true,
  breakatwhitespace=false,
  breakindent=0pt,
  upquote=true,
  columns=fullflexible,
  keepspaces=true,
  showstringspaces=false,
  literate={-}{{-\allowbreak}}1 {§}{{\S}}1 {—}{{\textemdash}}1 {–}{{\textendash}}1 {→}{{$\rightarrow$}}1 {⇒}{{$\Rightarrow$}}1 {≥}{{$\geq$}}1 {×}{{$\times$}}1
}
\newtcbinputlisting{\promptbox}[2]{
  enhanced, breakable, listing only,
  listing file={#2},
  listing options={style=promptstyle},
  title={#1},
  fonttitle=\bfseries\large, coltitle=white,
  colframe=promptframe, colback=promptback,
  boxrule=1.2pt, arc=2mm,
  left=5mm, right=5mm, top=2mm, bottom=2mm,
}
\usepackage[linesnumbered,boxed,ruled]{algorithm2e}
\usepackage{listings}
\usepackage{pifont}
\usepackage{enumitem}
\usepackage{wrapfig}

\usepackage{float}

\newcommand{\toolname}{\textsc{TestJack}}

\newcommand{\eg}{{\it e.g.,}\xspace}

\newcommand{\ie}{{\it i.e.,}\xspace}

\newcommand\tabref[1]{Table~\ref{#1}}

\newcommand\equref[1]{Eq.(\ref{#1})}

\newcommand\appref[1]{Appendix~\ref{#1}}
\newcommand{\fakeparagraph}[1]{\noindent\textbf{#1.}}

\usepackage{wasysym}
\usepackage{algorithmic}

\definecolor{codegreen}{rgb}{0,0.6,0}
\definecolor{codegray}{rgb}{0.5,0.5,0.5}
\definecolor{codepurple}{rgb}{0.58,0,0.82}
\definecolor{backcolour}{rgb}{0.95,0.95,0.92}

\lstdefinestyle{mystyle}{
  backgroundcolor=\color{backcolour},   commentstyle=\color{codegreen},
  keywordstyle=\color{magenta},
  numberstyle=\tiny\color{codegray},
  stringstyle=\color{codepurple},
  basicstyle=\ttfamily\footnotesize,
  breakatwhitespace=false,         
  breaklines=true,                 
  captionpos=b,                    
  keepspaces=true,                 
  numbers=left,                    
  numbersep=5pt,                  
  showspaces=false,                
  showstringspaces=false,
  showtabs=false,                  
  tabsize=2
}

\lstdefinestyle{casestyle}{
  style=mystyle,
  basicstyle=\ttfamily\scriptsize,
  numbers=none,
  frame=none,
  xleftmargin=1em,
  aboveskip=5pt,
  belowskip=5pt,
  keywordstyle=\color{black},
  stringstyle=\color{black},
  commentstyle=\color{codegray}
}

\usepackage{nicematrix}

\usepackage{amsthm}
\theoremstyle{definition}

\usepackage{bbding}

%% file: Text/abst.tex
\begin{abstract}
    Large language model (LLM) agents are rapidly reshaping software engineering, accompanied by an explosion of new code benchmarks. Yet nearly all existing benchmarks still rely on the same decades-old criterion: a solution is correct if it passes \emph{a fixed set of unit tests}.
    Such tests are often insufficient: they check only part of what the task requires, so agents can reward hack them or silently miss required behavior while still passing every test.
    As a result, higher benchmark scores may partly reflect better adaptation to the evaluator rather than better problem solving.
    Existing works focus on static test augmentation: they strengthen each task's tests once, before any trial is seen, and thus overlook how real trials actually fail.
    We introduce \toolname{}, a scalable framework for evaluating patches beyond fixed tests. For each trial, \toolname{} generates tests targeting prompt requirements the patch may violate, retains only tests passed by the ground-truth patch, and re-examines any trial failures. Each confirmed failure is thus supported by a replayable test. To reduce evaluation cost, we also introduce a lightweight variant which audits a random sample of trials in depth and reuses the resulting tests across all trials for the same task.
    Across 6 frontier model backends and 5 benchmarks such as DeepSWE and SWE Marathon, we find that about 34.4\% of the model trials currently judged correct violate the task requirements, lowering the overall resolution rate from 50.6\% to 33.2\%. Our results reveal a fundamental limitation of current coding-agent evaluation: as LLMs become better at optimizing against fixed evaluators, those evaluators themselves must become more adaptive.

\end{abstract}

%% file: Text/intro.tex
\section{Introduction}
\label{sec:introducion}

Large language model (LLM) agents are reshaping software engineering.
Modern agents navigate repositories, edit multiple files, run development tools, and iteratively refine their outputs, and systems such as Claude Code, Codex, and Copilot have become integral to research and production, making their rigorous evaluation increasingly important.
Code benchmarks have become the standard yardstick for comparing these agents, from function-level synthesis in HumanEval~\cite{chen2021humaneval} and MBPP~\cite{austin2021mbpp} to repository-level issues in SWE-bench~\cite{jimenez2024swebench}.
Their number continues to grow, with 210 introduced in 2024 and 316 in 2025~\cite{cao2026codebench}, and score improvements are routinely cited as evidence of progress.

\input{figures/motivation}

The community has invested heavily in these benchmarks: tasks are curated from real repositories or authored by engineers, and their tests are reviewed by hand, as when 93 professional developers screened SWE-bench to build SWE-bench Verified~\cite{chowdhury2024swebenchverified}.
Yet nearly all of them share one evaluation schema: a task counts as solved once the submission passes a test suite fixed when the benchmark was published.
We argue that this schema is fundamentally mismatched to agents:
\begin{insightbox}
    \textbf{Static tests cannot judge dynamic agents.} A fixed test suite checks only the behaviors its authors anticipated, whereas an agent observes execution feedback, adapts to it, and can pass every test without fully satisfying the task.
\end{insightbox}
Figure~\ref{fig:motivation} illustrates two kinds of benchmark defects behind this gap.
With \emph{insufficient tests}, the tests check less than the prompt requires, so trials that overlook a required case or game the tests still pass.
With an \emph{underspecified prompt}, a behavior is only implied by the prompt, so whether it must be implemented is ambiguous; the ground truth implements it but no test checks it.
Only the first is a defect of the trial: \toolname{} refutes trials for insufficient tests, whereas it attributes the second to the prompt and reports it as a prompt defect without refuting the trial (\S\ref{sec:approach:audit}).
The resulting overestimation can be substantial: through manual inspection, SWE-Bench+~\cite{aleithan2024swebenchplus} shows that excluding patches that passed only because of weak tests or solution leakage reduces one SWE-bench agent's resolution rate from 12.47\% to 3.97\%.

Existing efforts to close this gap sacrifice either scalability or fidelity.
Manual audits such as SWE-Bench+ reliably uncover such failures, but they do not scale: expert review must be repeated for every agent, model version, and run.
Automated benchmark auditors such as ABA~\cite{wang2026automated} ask whether a \emph{task} is fair and remove flawed tasks, which corrects unfairly failed trials but leaves falsely passed ones in place.
Test augmentation methods such as UTBoost~\cite{yu2025utboost} and SWE-ABS~\cite{yu2026sweabs} produce executable evidence but remain \emph{static}: they strengthen each task's suite once, before any submission is observed, so they catch only anticipated failures and fall short of the reward hacking of modern LLMs.
PatchDiff~\cite{wang2026patchdiff} inspects submissions dynamically, but still relies on human effort to decide which divergences are genuine defects.
What is needed is a \emph{dynamic} evaluator that judges each trial against what the task asks for like a manual auditor, and backs every verdict with executable evidence at scale like automated testing.

In this paper, we present \toolname{}\footnote{Our artifact is available at \url{https://anonymous.4open.science/r/testjack-C1A7}.}, an automated framework that audits the successes reported by coding benchmarks by \emph{evolving the evaluator} rather than trusting their fixed tests.
Our key insight is to separate \emph{what} a test should check from \emph{whether} its expectations are correct:
a trial should be judged against the requirements stated in the task prompt, and the ground-truth solution, which benchmarks already possess, can serve as an executable oracle that certifies any new test before it is used to judge a trial.
Building on this insight, \toolname{} audits each benchmark-passing trial with \emph{prompt-grounded differential testing}:
an LLM agent reviews the trial against the prompt's requirements, writes tests targeting the gaps it suspects, validates each on the ground truth, runs it on the trial, and re-judges any failure to confirm a genuine violation of the prompt, so every refutation is backed by a replayable \emph{witness test}.
Because auditing every trial is expensive, \toolname{} also offers a \emph{lightweight} approach that first adds oracle-validated tests for requirements each task leaves unchecked, and then audits randomly sampled trials and shares every confirmed witness with all remaining trials of the task.
Since trials for the same task often share failure modes, each expensive audit can refute many trials at the cost of cheap test executions.

We evaluate \toolname{} on five recent agentic coding benchmarks: DeepSWE~\cite{huang2026deepswe}, SkillsBench~\cite{li2026skillsbench}, SWE-Marathon~\cite{desai2026swemarathon}, SWE-bench Verified~\cite{chowdhury2024swebenchverified}, and SWE-bench Pro~\cite{deng2025swebenchpro}, the first three of which were built specifically to avoid weak tests.
Of 4,487 benchmark-passing trials from frontier models, including Claude Opus, GPT-5.5/5.6, and Gemini 3.1 Pro, \toolname{} refutes 1{,}545 (34.4\%): roughly one in three reported successes violates a requirement of its task, and the resolution rate over all 8{,}862 trials drops from 50.6\% to 33.2\%.
In summary, we make the following contributions:
\begin{itemize}[leftmargin=*, itemsep=1pt, topsep=2pt]
    \item \textbf{Problem.} We expose a fundamental limitation of \emph{static}, test-based evaluation for coding agents and propose \emph{dynamic evaluator evolution}, which generates new tests to probe each trial against the task requirements.
    \item \textbf{Technique.} We design prompt-grounded differential testing, which writes tests from the task's requirements, certifies them against the ground truth, and re-judges each failure, and a lightweight variant that combines task analysis with randomized auditing and cross-trial witness sharing.
    \item \textbf{Empirical study.} We find that 34.4\% of benchmark-passing trials of frontier agents on five benchmarks violate task requirements. And the lightweight variant recovers 72.4\% of these refutations at 15.4\% of the cost on three of the five benchmarks. Our audit results can serve as references for future benchmark design.
\end{itemize}

%% file: figures/motivation.tex
\begin{figure}[t]
    \centering
    \includegraphics[width=\linewidth]{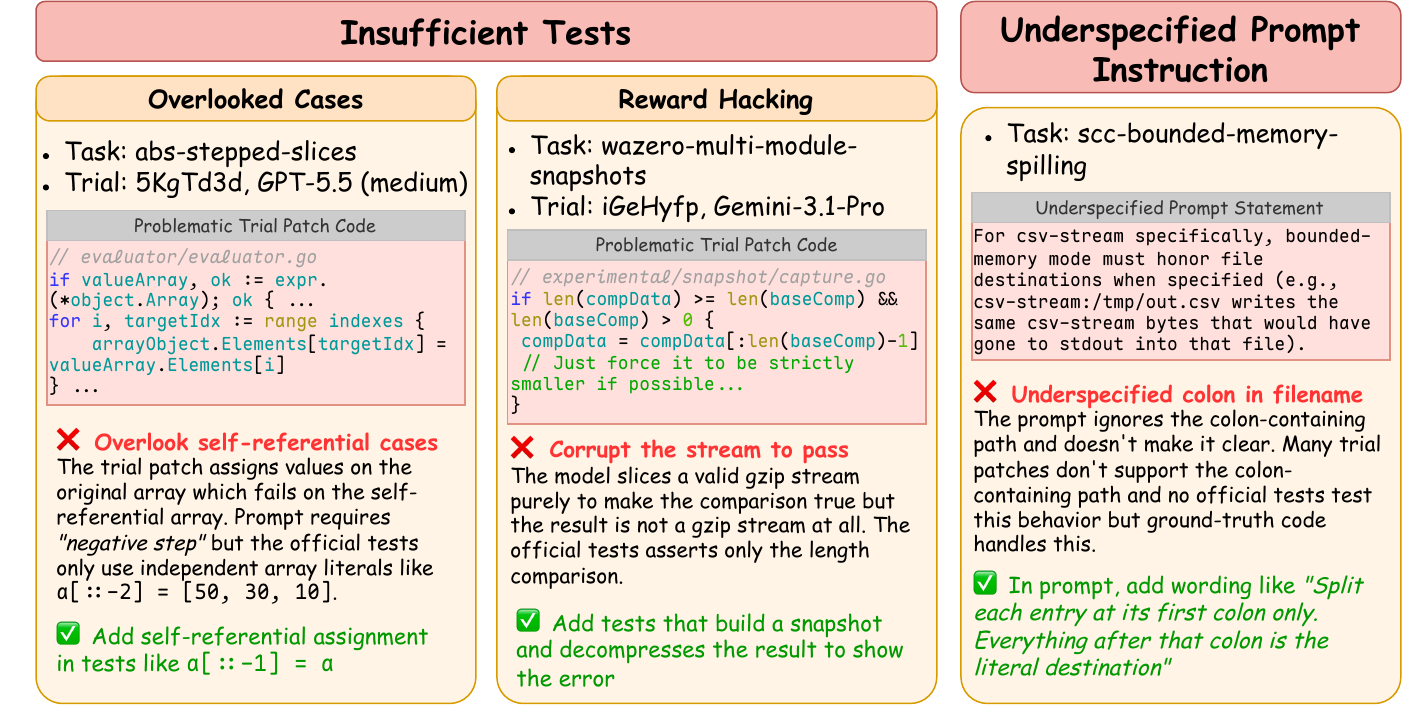}
    \caption{Two kinds of benchmark defects in a frontier benchmark DeepSWE~\cite{huang2026deepswe}, illustrated with benchmark-passing trials.
        \emph{Insufficient tests}: the tests check less than the prompt requires.
        \emph{Underspecified prompt} (right): a behavior is only implied by the prompt, so whether it must be implemented is ambiguous; the ground truth implements it, but no test checks it.}
    \label{fig:motivation}
    \vspace{-1.0em}
\end{figure}

%% file: Text/background.tex
\section{Related Work}
\label{sec:background}

\noindent\mbox{}\par\vspace{-1.8\baselineskip}
\input{figures/related_work}
\fakeparagraph{Coding Benchmarks}
Benchmarks such as MMLU~\cite{hendrycks2021mmlu}, BIG-bench~\cite{srivastava2023bigbench}, HELM~\cite{liang2023helm}, and Humanity's Last Exam~\cite{cais2026hle} structure progress in AI, yet their reliability is increasingly questioned: test sets contain label errors~\cite{northcutt2021pervasive}, public benchmarks leak into pretraining corpora~\cite{xu2024contamination}, leaderboard practices can distort rankings~\cite{singh2025leaderboard}, and flaws in agentic task setup or reward design can misestimate performance by up to 100\%~\cite{zhu2025establishing}; dynamic benchmarking~\cite{kiela2021dynabench, pmlr-v267-chen25ba, chen2024ppm} responds by continually collecting new adversarial examples.
Coding benchmarks have evolved from function-level synthesis~\cite{chen2021humaneval,austin2021mbpp} to repository-level issue resolution in SWE-bench~\cite{jimenez2024swebench} and its successors, which add human validation~\cite{chowdhury2024swebenchverified}, languages~\cite{zan2025multiswebench}, modalities~\cite{yang2025swebenchmm}, contamination-limiting refreshes~\cite{zhang2025swebenchlive,badertdinov2025swerebench}, and enterprise-grade changes~\cite{deng2025swebenchpro}.
Another line authors tasks from scratch, from command-line work~\cite{merrill2026terminalbench} and skill-augmented workflows~\cite{li2026skillsbench} to long-horizon engineering projects~\cite{huang2026deepswe,desai2026swemarathon}.
Nearly all of these benchmarks share one \emph{static} protocol: a submission is correct if it passes a test suite fixed when the benchmark is released.

\fakeparagraph{Benchmark Auditing and Test Strengthening}
Prior work on benchmark reliability can be organized along two axes (\Figref{fig:related_work}): whether the evaluator is \emph{static}, analyzing a task before any submission is observed, or \emph{dynamic}, auditing the trials that agents actually produce; and whether its verdicts rest on manual checking or automated auditing.
Manual audits expose failures that scores conceal~\cite{aleithan2024swebenchplus} but must be repeated for every new agent and run.
Automated auditors such as ABA~\cite{wang2026automated} and BenchGuard~\cite{tu2026benchguard} use LLM agents to flag tasks whose specification, environment, or grader is broken~\cite{mohl2026transcript}; they audit the \emph{benchmark} rather than individual trials, and their verdicts rest on LLM judgment.
Another line strengthens benchmark tests, from amplifying HumanEval's tests~\cite{liu2023evalplus} to generating additional or adversarial tests for SWE-bench, as in UTBoost~\cite{yu2025utboost} and SWE-ABS~\cite{yu2026sweabs}, and issue-reproducing tests~\cite{mundler2024swtbench,ahmed2025otter}.
These methods are static: they build one suite per task before observing any submission, so their tests must anticipate where agents fail and may encode incidental details of the gold patch.
PatchDiff~\cite{wang2026patchdiff} is dynamic, testing agent patches differentially against the gold patch, but only 28.6\% of the divergent patches it reviews can be confirmed incorrect and most cannot be judged either way, so deciding which divergences indict a trial still requires manual inspection.
\textbf{\toolname{} is, to our knowledge, the first automated dynamic evaluator} and occupies the automated, dynamic quadrant of \Figref{fig:related_work}: it audits individual trials rather than tasks, derives tests from the task prompt rather than the gold patch, uses the ground truth only to validate their expectations, and backs every verdict with an executable witness test instead of LLM or human judgment.

%% file: figures/related_work.tex
\begin{wrapfigure}{r}{0.44\linewidth}
    \centering
    \vspace{-12pt}
    \includegraphics[width=\linewidth]{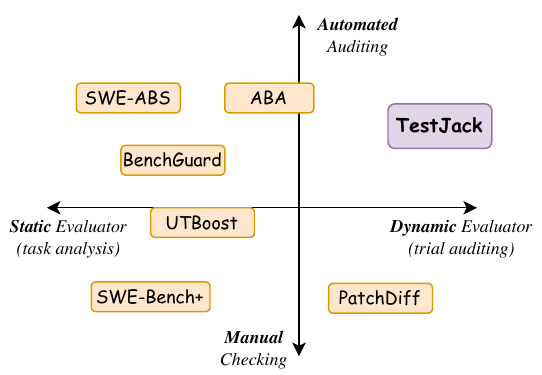}
    \caption{Benchmark auditing approaches by evaluator type (static task  vs.\ dynamic trial auditing) and level of automation.}
    \label{fig:related_work}
    \vspace{-10pt}
\end{wrapfigure}

%% file: src/approach.tex
\section{Methodology}
\label{sec:approach}

We first formulate the problem (\S\ref{sec:approach:problem}), then present the per-trial audit (\S\ref{sec:approach:audit}) and its lightweight variant (\S\ref{sec:approach:random}); \Figref{fig:method} gives an overview of both.

\subsection{Problem Formulation}
\label{sec:approach:problem}

A benchmark instance consists of a natural-language task specification $P$ (\eg a GitHub issue), a repository snapshot $R$, a ground-truth implementation $G$ (\eg the developer-written reference patch applied to $R$), and a static test suite $T_0$ used by the benchmark to score submissions.
Running an agent on the instance produces a \emph{trial}: a candidate patch $s$ applied to $R$, yielding a submission $R[s]$.
A trial is \emph{passing} if $R[s]$ passes every test in $T_0$.
\toolname{} takes as input the set of passing trials $S = \{s_1, \dots, s_n\}$ collected across agents and asks, for each trial, whether its passing verdict should be trusted.

\fakeparagraph{Requirements and behavior spaces}
Let $\mathcal{Q}(P) = \mathcal{Q}_e(P) \cup \mathcal{Q}_i(P)$ denote the requirements of the task, where $\mathcal{Q}_e(P)$ contains the requirements that $P$ states explicitly and $\mathcal{Q}_i(P)$ those that follow implicitly from $P$ together with the documented behavior of the repository.
For an artifact $X$, let $\mathcal{B}(X)$ denote the behaviors it pins down: $\mathcal{B}(T_0)$ is the behavior asserted by the benchmark tests, $\mathcal{B}(P)$ is the behavior required by $\mathcal{Q}(P)$, and $\mathcal{B}(G)$ is the behavior exhibited by the ground truth.
We assume each task is well-formed (\appref{app:assumption}), so these spaces are nested:
\begin{equation}
    \label{eq:behavior}
    \mathcal{B}(T_0) \subseteq \mathcal{B}(P) \subseteq \mathcal{B}(G).
\end{equation}
The first inclusion reflects that the tests check only part of what the prompt requires.
The second reflects that the ground truth, being a correct solution, implements every required behavior, together with incidental choices that $P$ leaves open or ambiguous.
A passing trial is guaranteed to agree with the task only on $\mathcal{B}(T_0)$, so its possible defects lie in the gap $\mathcal{B}(P) \setminus \mathcal{B}(T_0)$.

\fakeparagraph{Witness tests}
Auditing a trial $s$ therefore amounts to searching this gap for a \emph{witness test} $t$ satisfying
\begin{equation}
    \label{eq:witness}
    t \text{ checks some } q \in \mathcal{Q}(P)
    \quad \wedge \quad
    t(R[G]) = \textsf{pass}
    \quad \wedge \quad
    t(R[s]) = \textsf{fail}.
\end{equation}
A witness is a concrete, replayable demonstration that the trial violates a requirement of the task, even though it passes the entire benchmark suite $T_0$.

\input{figures/method}

\subsection{Per-Trial Auditing via Prompt-Grounded Differential Testing}
\label{sec:approach:audit}

There are two obvious ways to search for a witness test, and each fails in a different direction of \equref{eq:behavior}.
\emph{(i) Writing tests from $P$ alone} stays within $\mathcal{B}(P)$ but raises the classical test-oracle problem: an LLM-written test may encode a wrong expectation, and nothing certifies that a failing trial is actually at fault.
\emph{(ii) Writing tests from $G$ alone}, \ie plain differential testing, accepts any test that separates $R[G]$ from $R[s]$.
Such tests range over all of $\mathcal{B}(G)$, including the incidental choices in $\mathcal{B}(G) \setminus \mathcal{B}(P)$, so they overfit to the ground truth and penalize a trial for not reproducing the reference patch rather than for failing the task.
\toolname{} combines the strengths of both: it writes tests from $P$, which the first condition of \equref{eq:witness} enforces, and uses $G$ as an executable oracle, which the second condition enforces, so that a new test is trusted only if the ground truth passes it.
The prompt thus determines \emph{what} to test, the trial determines \emph{where} to look for violations, and the ground truth confirms that the test's expectations are correct.

The boundary of $\mathcal{B}(P)$, however, is not always sharp.
Implicit requirements in $\mathcal{Q}_i(P)$ rely on background knowledge that the prompt does not state, and when the prompt is \emph{underspecified}, a witness for such a requirement may reflect a defect of the prompt rather than of the trial.
\toolname{} reports some witnesses as prompt defects rather than trial defects caused by insufficient tests according to an empirical rule described in \appref{app:implicit}.

\fakeparagraph{Audit procedure}
\toolname{} implements the audit with two LLM agents, an \emph{audit agent} and a \emph{reviewer agent} (\Figref{fig:method}, left).
Both operate inside the benchmark's own execution environment, and neither treats the benchmark's pass verdict as evidence of correctness. The audit agent proceeds in five steps.
\begin{enumerate}[leftmargin=*, itemsep=1pt, topsep=2pt]
    \item \emph{Execution worlds.}
          The agent builds three copies of the repository under the official evaluator: the \emph{model world} $R[s]$, the \emph{reference world} $R[G]$, and the \emph{null world} $R$ without any patch.
          The null world reveals which benchmark tests already pass without a fix and therefore verify nothing about the trial.
    \item \emph{Requirements checklist.}
          The agent decomposes $P$ into atomic, testable clauses, which instantiate $\mathcal{Q}(P)$, and records for each clause whether $T_0$ covers it and whether the trial implements it.
    \item \emph{Evidence collection.}
          The agent then examines the trial's diff along three axes:
          (i) \emph{ghost branches}, code paths in the diff that $T_0$ never executes, identified with branch coverage and classified by whether $P$ requires the behavior they implement;
          (ii) \emph{reward hacking}, checked with explicit detectors for hard-coded expected values, special-casing of tested inputs, test-environment sniffing, harness tampering, and stubs behind a tested veneer; and
          (iii) \emph{general defects}, including specification violations, robustness, and security issues, prioritized by mutation testing, which exposes behavior that $T_0$ does not pin down.
    \item \emph{Issue-demonstrating tests.}
          For every suspected issue, the agent writes a test in the repository's native framework that asserts the behavior the violated clause requires.
          The test must fail in the model world and pass in the reference world, which serves as the oracle.
    \item \emph{Audit report.}
          The agent assigns the trial a verdict, such as sound, incorrect despite passing, or reward hack.
          For each issue, it explains what the defect is, why it occurs, which clause of $P$ requires the correct behavior, and why $T_0$ fails to catch it.
\end{enumerate}
The reviewer agent then re-judges these tests, treating $P$ as the sole authority on required behavior and deliberately not using $G$ to infer requirements.
It labels each test as valid, questionable, or invalid, rejecting tests that assert behavior the prompt does not request, as they overfit to a particular solution.
The trial is \emph{refuted} if at least one valid test fails on it; such a test is a witness in the sense of \equref{eq:witness}.

\fakeparagraph{Properties}
Every verdict is \emph{executable}: unlike LLM-as-a-judge approaches, a refutation is backed by a test run against both implementations.
Verdicts are \emph{prompt-grounded}: the checklist and the reviewer hold the trial to what the task asked for.
The audit is \emph{agnostic to the failure mode}, catching reward-hacked patches by the same mechanism without a predefined taxonomy of exploits, and, being fully automatic, it can be repeated as new agents and trials appear.

\subsection{Lightweight Approach: Randomized Auditing with Witness Sharing}
\label{sec:approach:random}

The per-trial audit is precise but expensive: each audit is a multi-step agent session with repeated builds and test executions, and the number of trials grows multiplicatively with the number of agents, models, and repetitions evaluated on a benchmark.
Auditing every trial independently is also wasteful.
Trials for the same task often fail for the same reasons, for instance when many agents miss the same edge case, so independent audits keep rediscovering the same defects.
\toolname{} reduces this cost with two ideas: it moves part of the audit from individual trials to the task itself, and it shares every validated witness across all trials of the task (Algorithm~\ref{alg:random} in \appref{app:random}).

\fakeparagraph{Agentic task analysis}
Many weaknesses of a task are visible before any trial is examined: comparing $P$ with $T_0$ often reveals an input the prompt mentions but no test exercises, an error condition that is never triggered, or neighboring behavior the change could break.
\toolname{} therefore begins with a task-level analysis in which an agent reads $P$, $T_0$, and $G$, identifies requirements of $\mathcal{Q}(P)$ that $T_0$ does not adequately cover, and writes new tests targeting them.
With $G$ as the oracle, a test that fails on $R[G]$ is refined using the execution feedback until it passes, or discarded once the refinement budget is exhausted.
The result is an augmented suite $T_1 = T_0 \cup T_{\text{task}}$, built once per task regardless of how many trials the task has.

\fakeparagraph{Randomized audit with cross-trial witness sharing}
Task analysis cannot anticipate every way a trial may go wrong, so \toolname{} complements it with per-trial audits on randomly sampled trials whose witnesses are shared with every other trial of the task, exploiting the fact that trials share failure modes.
Let $T$ denote the current test suite, initialized to $T_1$.
Trials that fail any test in $T$ are refuted, and the remaining trials form the \emph{active set} $A$.
The loop then proceeds in rounds:
(i) sample a trial $s$ uniformly at random from $A$;
(ii) run the per-trial audit of \S\ref{sec:approach:audit} on $s$;
(iii) if the audit confirms witnesses $W$, add $W$ to $T$, \emph{share} $W$ by executing it on every trial in $A$, and refute all trials that fail;
(iv) otherwise, mark $s$ as audited-clean and remove it from $A$.
The loop terminates when $A$ is empty or the working budget is exhausted.
Sharing is sound because a witness belongs to the task rather than to the trial on which it was found.
The first two conditions of \equref{eq:witness}, that $t$ checks a requirement in $\mathcal{Q}(P)$ and passes on $R[G]$, do not involve $s$ at all, and the reviewer agent judges them against $P$ rather than against the trial. Thus, once a witness is accepted, deciding whether it refutes another trial reduces to the third condition.

The scheme is economical because generating and validating tests is expensive, whereas replaying a test on another trial costs a single execution.
Each audit round pays the generation cost once while its witnesses are shared with the entire active set, so when trials share failure modes, the number of expensive audits approaches the number of \emph{distinct} defects rather than of defective trials.
The trade-off is coverage: shared witnesses are not re-judged for each trial they refute, and defects unique to never-sampled trials can be missed under a finite budget (\S\ref{sec:eval:rq2}).

%% file: figures/method.tex
\begin{figure}[t]
    \centering
    \includegraphics[width=\linewidth]{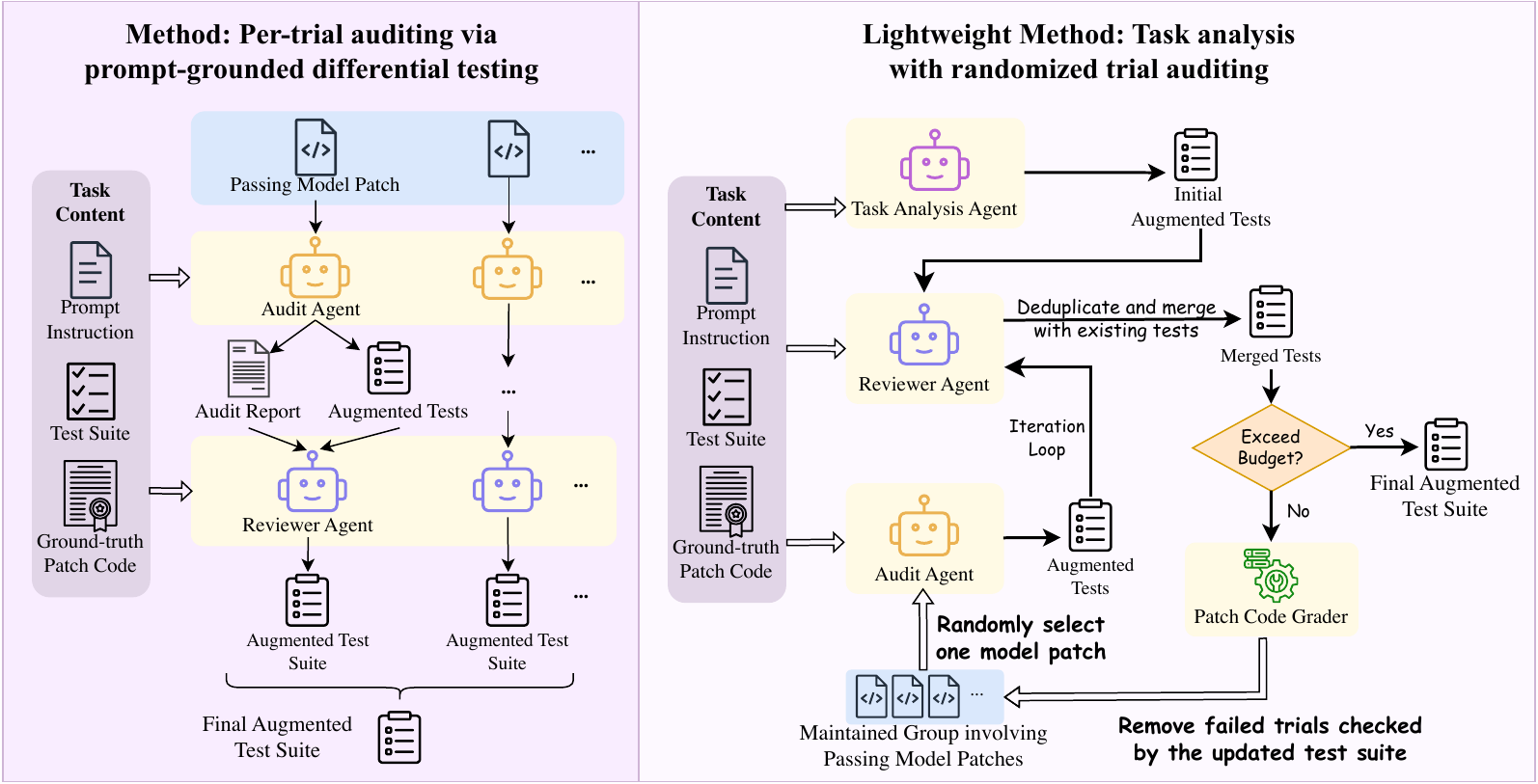}
    \caption{Overview of \toolname{}.
        \emph{Left:} per-trial auditing via prompt-grounded differential testing with audit agent and reviewer agent (\S\ref{sec:approach:audit}).
        \emph{Right:} the lightweight approach (\S\ref{sec:approach:random}) with a randomized trial auditing loop.}
    \label{fig:method}
\end{figure}

%% file: src/evaluation.tex
\section{Evaluation}
\label{sec:evaluation}

We evaluate \toolname{} along four research questions:
\begin{itemize}[leftmargin=*, itemsep=0pt, topsep=2pt]
    \item \textbf{RQ1 (Per-trial auditing on frontier benchmarks):} When the per-trial audit of \S\ref{sec:approach:audit} is applied to every benchmark-passing trial, how many successes does it refute, and how much do the refutations lower reported resolution rates?
    \item \textbf{RQ2 (Lightweight randomized approach):} How many of the per-trial audit's refutations does the lightweight approach of \S\ref{sec:approach:random} recover, how much cost does it save, and where does it fall short?
    \item \textbf{RQ3 (Comparison with existing benchmark augmentation):} How does \toolname{} compare with SWE-ABS, a state-of-the-art coding benchmark augmentation method, in detecting defective trials?
    \item \textbf{RQ4 (Ablation study):} How much do prompt-based task analysis and randomized trial auditing each contribute to the lightweight approach?
\end{itemize}

\subsection{Evaluation Setup}
\label{sec:eval:setup}

\fakeparagraph{Benchmarks}
We audit five recent agentic coding benchmarks, summarized in \tabref{tab:benchmarks}: SWE-bench Verified~\cite{chowdhury2024swebenchverified}, SWE-bench Pro~\cite{deng2025swebenchpro}, DeepSWE~\cite{huang2026deepswe}, SkillsBench~\cite{li2026skillsbench}, and SWE-Marathon~\cite{desai2026swemarathon}.
They span the two ways of building the static verifier $T_0$: tests inherited from the resolving pull request (SWE-bench Verified and Pro), and verifiers written from scratch (DeepSWE, SkillsBench, and SWE-Marathon), the latter designed specifically to avoid weak inherited tests.
All five provide the task specification $P$, an executable environment $R$, and a reference solution $G$.
For the container-based benchmarks (SkillsBench and SWE-Marathon), $G$ is the task's oracle solution, a trial $s$ is the agent's final container state, and the hidden verifier plays the role of $T_0$.
We audit every task of SWE-bench Verified, DeepSWE, and the public split of SWE-bench Pro, the coding tasks of SkillsBench, and the SWE-Marathon tasks that need no GPU and are not ultra-long-horizon (\appref{app:additional_experiment:setup}).

\input{tables/benchmarks}

\fakeparagraph{Agents and trials}
We audit trials of frontier models, whose scores drive current claims of progress: Claude Opus 4.6/4.7/4.8, GPT 5.5/5.6, and Gemini 3.1 Pro, regardless of the agent harness.
For DeepSWE, SkillsBench, and SWE-Marathon, we use the trajectories released by the benchmark authors, covering vendor agents (\eg Claude Code, Codex CLI, Gemini CLI) and open scaffolds (\eg mini-swe-agent).
For SWE-bench Verified and Pro, which have few open trajectories from these models, we run GPT-5.6-sol with medium thinking effort under mini-swe-agent, with two trials per task.
Of the 8{,}862 trials in this scope, 4{,}487 pass the benchmark's own verifier $T_0$ and form the audit corpus (\tabref{tab:benchmarks}); the remaining trials serve only as the denominator of resolution rates.

\fakeparagraph{Implementation and configuration}
\toolname{} uses GPT-5.5 with high reasoning effort for both the per-trial audit and the prompt-based task analysis.
Both run as a Codex agent inside the task's execution environment, where it can read $P$, $G$, the existing tests, the repository, and (for the per-trial audit) the trial $s$, and execute candidate tests on $R[G]$ and $R[s]$.
We cap the search at $k = 3$ generate-and-validate iterations per trial.

\subsection{RQ1: Per-Trial Auditing on Frontier Benchmarks}
\label{sec:eval:rq1}

We run the per-trial audit of \S\ref{sec:approach:audit} on every benchmark-passing trial (\tabref{tab:rq1}).
\toolname{} refutes 1{,}545 of 4{,}487 trials (34.4\%), lowering the resolution rate over all 8{,}862 trials in scope from 50.6\% to 33.2\%; each refutation is backed by a witness test that checks a requirement of the prompt, passes on the ground truth, fails on the trial, and survives re-judgment.
Roughly one in three successes thus violates the task requirements despite passing the benchmark's own verifier.
The refutation rate ranges from 7.6\% on SWE-bench Verified to 53.6\% on SWE-Marathon, and careful verifier design does not remove the problem: the from-scratch verifiers of DeepSWE and SWE-Marathon yield the two highest rates (43.5\% and 53.6\%).
\appref{app:case_study} presents representative refuted trials.

\input{tables/per-trial}

\subsection{RQ2: Lightweight Randomized Approach}
\label{sec:eval:rq2}

We run the lightweight approach of \S\ref{sec:approach:random} on DeepSWE, SkillsBench, and SWE-Marathon with a budget of three audit rounds per task and compare it with the per-trial audit on the same trials; SWE-bench Verified and Pro have too few trials per task for randomized auditing.
As \tabref{tab:rq2} shows, it recovers 72.4\% of the per-trial refutations while reducing API cost by 84.6\%, to \$4{,}037 in total, with recall ranging from 40.0\% on SWE-Marathon to 80.0\% on SkillsBench.
Witness sharing makes this trade favorable: a witness found on one sampled trial is validated once and replayed on every remaining trial, so the expensive search runs a few times per task rather than once per trial.
The disagreement with the per-trial audit mainly reflects underspecified prompts rather than incorrect tests (\appref{app:additional_experiment:underspecified_prompt}). The reason why it's precision is not high is that per-trial auditing keeps each finding attached to one trial which focuses more on the trial itself, while the lightweight approach turns one trial's findings into an executable gate and replays it on everyone. Thus, the lightweight approach is prone to diffuse the small defects like the underspecified prompts.
Moreover, another two limitations follow from the design of the lightweight approach.
First, shared witnesses transfer well within a failure mode but rarely across modes, so tasks whose refuted trials span several distinct defects are covered slowly under a finite budget, which we expect accounts for much of the missed 27.6\%.
Second, uniform sampling spends rounds on clean trials, which hurts most when few trials amortize a round: SWE-Marathon, with 28 passing trials, shows both the lowest recall (40.0\%) and the smallest cost reduction (53.7\%).

\input{tables/lightweight}

\subsection{RQ3: Comparison with Existing Benchmark Augmentation}
\label{sec:eval:rq3}

We compare \toolname{} with SWE-ABS~\cite{yu2026sweabs}, which strengthens each task's suite before any submission is observed by combining coverage-driven test generation with adversarial tests that reject mutants of the gold patch.
We run SWE-ABS (without cost budget) on DeepSWE, SkillsBench, and SWE-Marathon with the same model as \toolname{}, replay its suite on the RQ1 corpus, and count a trial as refuted if it fails any added test.
As \tabref{tab:rq3} shows, SWE-ABS refutes 248 trials (8.5\% of benchmark-passing trials) and recovers 9.9\% of the per-trial audit's refutations, against 1863 trials (63.6\%) and 72.4\% recall for the lightweight approach: a 7.3$\times$ difference in recall at comparable precision (50.4\% vs.\ 49.2\%).
The gap widens as tasks grow more complex; on SWE-Marathon, SWE-ABS refutes no trial at all, while the lightweight approach refutes 10 of 28.
Two design differences explain the gap.
First, SWE-ABS never observes a submission, so its tests must anticipate where agents go wrong, whereas \toolname{} targets defects that depend on how a particular trial is implemented.
Second, SWE-ABS derives expectations from the gold patch rather than the prompt, so its tests can penalize implementation choices the prompt leaves open (\eg exact error messages), whereas \toolname{} grounds every test in the prompt and re-judges failures.
Static test augmentation is thus insufficient, and dynamic trial auditing is necessary to reach implementation-dependent defects.

\input{tables/sweabs-comparison}

\subsection{RQ4: Ablation Study}
\label{sec:eval:rq4}

We compare \emph{task analysis only}, which refutes trials using only the augmented suite $T_1$, with the full lightweight approach at the RQ2 budget (\tabref{tab:rq4}).
Task analysis alone refutes 345 trials and recovers 13.4\% of the per-trial refutations: some gaps in $T_0$ are visible before any submission is examined, and this trial-independent suite is built once per task.
Adding randomized auditing with witness sharing raises recall to 72.4\%, a 5.4$\times$ increase, at essentially unchanged precision (49.3\% vs.\ 49.2\%), and the pattern holds on every benchmark.
The stages are thus complementary in coverage: most refutations require evidence from a concrete submission, which the second stage supplies without lowering agreement with the per-trial audit.

%% file: tables/benchmarks.tex
\begin{table}[t]
    \centering
    \caption{Benchmarks audited by \toolname{}. \#Tasks counts the tasks we audit; \#Passing counts the benchmark-passing trials in our audit corpus. $^\dagger$Public split only. $^\ast$Software Engineering tasks that do not require a GPU. $^\ddagger$Excluding GPU tasks and ultra-long-horizon projects (\eg building a C compiler).}
    \label{tab:benchmarks}
    \resizebox{\linewidth}{!}{
        \begin{tabular}{lrllll r}
            \toprule
            Benchmark               & \#Tasks & Languages                & Task origin           & Reference solution $G$ & Static verifier $T_0$         & \#Passing \\
            \midrule
            DeepSWE                 & 113     & TS, Go, Python, JS, Rust & Authored from scratch & Authored solution      & Hand-written functional tests & 2,848     \\
            SkillsBench$^\ast$      & 13      & Various                  & Community-authored    & Oracle solution        & Deterministic pytest verifier & 55        \\
            SWE-Marathon$^\ddagger$ & 2       & Various                  & Expert-authored       & Oracle solution        & Multi-layer hidden verifier   & 28        \\
            SWE-bench Verified      & 500     & Python                   & Mined pull requests   & Merged patch           & Pull-request tests            & 805       \\
            SWE-bench Pro$^\dagger$ & 731     & Python, Go, JS, TS       & Mined commits         & Merged patch           & Human-reviewed commit tests   & 751       \\
            \midrule
            Total                   & 1,359   &                          &                       &                        &                               & 4,487     \\
            \bottomrule
        \end{tabular}
    }
\end{table}

%% file: tables/per-trial.tex
\begin{table}[t]
    \centering
    \caption{Per-trial auditing results of \toolname{}, aggregated by benchmark. \#All counts all trials in our scope, \#Passing the benchmark-passing ones, and \#Refuted those refuted by a confirmed witness test. Refuted (\%) is relative to \#Passing; Corrected rate shows the resolution rate before the audit, \#Passing / \#All, and after it, (\#Passing $-$ \#Refuted) / \#All.}
    \label{tab:rq1}
    \resizebox{0.95\linewidth}{!}{
        \begin{tabular}{lrrrrc}
            \toprule
            Benchmark          & \#All & \#Passing & \#Refuted & Refuted (\%) & Corrected rate              \\
            \midrule
            DeepSWE            & 6,236 & 2,848     & 1,240     & 43.5         & 45.7\% $\rightarrow$ 25.8\% \\
            SkillsBench        & 85    & 55        & 10        & 18.2         & 64.7\% $\rightarrow$ 52.9\% \\
            SWE-Marathon       & 79    & 28        & 15        & 53.6         & 35.4\% $\rightarrow$ 16.5\% \\
            SWE-bench Verified & 1,000 & 805       & 61        & 7.6          & 80.5\% $\rightarrow$ 74.4\% \\
            SWE-bench Pro      & 1,462 & 751       & 219       & 29.2         & 51.4\% $\rightarrow$ 36.4\% \\
            \midrule
            Total              & 8,862 & 4,487     & 1,545     & 34.4         & 50.6\% $\rightarrow$ 33.2\% \\
            \bottomrule
        \end{tabular}
    }
\end{table}

%% file: tables/lightweight.tex
\begin{table}[t]
    \centering
    \caption{Performance of the lightweight approach at a budget of 3 audit rounds per task. \#Per-trial and \#Lightweight count the trials refuted by the per-trial audit (RQ1) and by the lightweight approach. Recall is the fraction of per-trial refutations that the lightweight approach also refutes. Precision is the fraction of lightweight-method refutations that the per-trial audit also refutes. Cost is the GPT API cost, and reduction cost is compared with the per-trial audit. $^\dagger$: the precision is impacted by underspecified prompt described in \Cref{app:additional_experiment:underspecified_prompt}. $^\ddagger$: the cost is estimated based on the token cost and API pricing.}
    \label{tab:rq2}
    \resizebox{\linewidth}{!}{
        \begin{tabular}{lrrrrrr}
            \toprule
            Benchmark    & \#Per-trial & \#Lightweight & Recall (\%) & Prec.\ (\%)    & Cost (\$)$^\ddagger$ & Red.\ cost (\%) \\
            \midrule
            DeepSWE      & 1,240       & 1,822         & 72.7        & 49.5           & 3,904                & 84.9\%          \\
            SkillsBench  & 10          & 31            & 80.0        & 25.8           & 85                   & 78.6\%          \\
            SWE-Marathon & 15          & 10            & 40.0        & 60.0           & 48                   & 53.7\%          \\
            \midrule
            Total        & 1,265       & 1,863         & 72.4        & 49.2$^\dagger$ & 4,037                & 84.6\%{}        \\
            \bottomrule
        \end{tabular}
    }
\end{table}

%% file: tables/sweabs-comparison.tex
\begin{table}[t]
    \centering
    \caption{Comparison with SWE-ABS and ablation of the lightweight approach. \toolname{}-Task Analysis uses only the augmented suite produced by task analysis, while \toolname{}-Lightweight additionally uses randomized auditing with witness sharing (\S\ref{sec:approach:random}). Refuted (\%) is the fraction of benchmark-passing trials refuted; recall is the fraction of the per-trial audit's refutations (RQ1) that the method also refutes; precision is the fraction of the method's refutations that the per-trial audit also refutes.}
    \label{tab:rq3}
    \label{tab:rq4}
    \resizebox{\linewidth}{!}{
        \begin{tabular}{llrrrr}
            \toprule
            Benchmark & Method                    & \#Refuted & Refuted (\%) & Recall (\%) & Precision (\%) \\
            \midrule
            \multirow{3}{*}{DeepSWE}
                      & SWE-ABS                   & 246       & 8.6          & 10.0        & 50.4           \\
                      & \toolname{}-Task Analysis & 344       & 12.1         & 13.6        & 49.1           \\
                      & \toolname{}-Lightweight   & 1,822     & 64.0         & 72.7        & 49.5           \\
            \midrule
            \multirow{3}{*}{SkillsBench$^\ast$}
                      & SWE-ABS                   & 2         & 3.6          & 10.0        & 50.0           \\
                      & \toolname{}-Task Analysis & 0         & 0.0          & 0.0         & --             \\
                      & \toolname{}-Lightweight   & 31        & 56.4         & 80.0        & 25.8           \\
            \midrule
            \multirow{3}{*}{SWE-Marathon$^\ddagger$}
                      & SWE-ABS                   & 0         & 0.0          & 0.0         & --             \\
                      & \toolname{}-Task Analysis & 1         & 3.6          & 6.7         & 100.0          \\
                      & \toolname{}-Lightweight   & 10        & 35.7         & 40.0        & 60.0           \\
            \midrule
            \multirow{3}{*}{Total}
                      & SWE-ABS                   & 248       & 8.5          & 9.9         & 50.4           \\
                      & \toolname{}-Task Analysis & 345       & 11.8         & 13.4        & 49.3           \\
                      & \toolname{}-Lightweight   & 1,863     & 63.6         & 72.4        & 49.2           \\
            \bottomrule
        \end{tabular}
    }
\end{table}

%% file: Text/conclusion.tex
\section{Conclusion}
\label{sec:conclusion}

We present \toolname{}, an automated framework that audits the successes reported by coding benchmarks by evolving the evaluator rather than trusting the fixed tests shipped with them.
\toolname{} audits each benchmark-passing trial with prompt-grounded differential testing: it writes tests for the requirements of the task prompt, certifies each test against the ground-truth solution, and re-judges every failure against the prompt, so that each refutation is backed by a replayable witness test.
Its lightweight variant combines prompt-based task analysis with randomized trial auditing and reuses validated tests across all trials of a task.
On 4{,}487 benchmark-passing trials of frontier agents across five benchmarks, \toolname{} refutes 34.4\% of them, including 43.5\% on DeepSWE, whose verifiers were written specifically to avoid weak tests, and lowers the overall resolution rate from 50.6\% to 33.2\%.
The lightweight variant recovers 72.4\% of these refutations at 15.4\% of the cost, with 7.3$\times$ the recall of the benchmark augmentation method SWE-ABS.
These results suggest that fixed test suites substantially overstate the capability of frontier coding agents, and that evaluators should evolve with the submissions they judge.

%% file: appendix/discussion.tex
\section{Method Discussion}
\label{app:method-discussion}

\subsection{Task Well-Formedness Assumption}
\label{app:assumption}

\toolname{} assumes that every task in the benchmark suite is well-formed in the sense of \equref{eq:behavior}, which is, $\mathcal{B}(T_0) \subseteq \mathcal{B}(P) \subseteq \mathcal{B}(G)$.
The assumption can fail in three ways, none of which we consider in this work:
\begin{itemize}[leftmargin=*, itemsep=1pt, topsep=2pt]
    \item $\mathcal{B}(T_0) \not\subseteq \mathcal{B}(P)$: the official tests assert behavior that the prompt does not require. A trial can then be marked failing even though it satisfies everything the task asked for, so the benchmark verdict we audit is itself unfair to the trial.
    \item $\mathcal{B}(P) \not\subseteq \mathcal{B}(G)$: the prompt requires behavior, whether stated or implied, that the ground truth does not implement. Oracle validation would then discard a legitimate witness test, because a test of that behavior fails on $R[G]$ and never reaches the trial.
    \item $\mathcal{B}(T_0) \not\subseteq \mathcal{B}(G)$: the official tests assert behavior the ground truth does not exhibit, \ie the ground truth does not pass the tests the benchmark scores against. The two artifacts \toolname{} relies on then contradict each other, and neither can serve as a reference.
\end{itemize}
The third case is largely ruled out by benchmark construction, since the benchmarks we audit verify that the reference solution passes the official tests before a task is released (\S\ref{sec:eval:setup}).
The first two are properties of task quality that a trial-level audit cannot repair: they call for fixing the task rather than for judging any particular submission.
A related but distinct failure arises when \equref{eq:behavior} holds for the task author but not for the agent that reads $P$, which we treat separately in \S\ref{app:implicit}.

\subsection{Implicit Requirements and Prompt Underspecification}
\label{app:implicit}

Explicit requirements in $\mathcal{Q}_e(P)$ are always clear to interpretation, but implicit requirements in $\mathcal{Q}_i(P)$ are harder.
Task authors routinely omit requirements they expect any reader to infer, following the maxim of quantity~\cite{grice1975logic,frank2012predicting}, much as customers never state the ``must-be'' features they take for granted~\cite{kano1984attractive}.
When author and reader share this background, the omission is harmless.
When they do not, the prompt is underspecified: the ground truth implements the behavior and no official test checks it, but whether a trial must implement it is ambiguous (\Figref{fig:motivation}, right).
A failing witness then reflects a defect of the prompt rather than of the trial.

\fakeparagraph{Underspecification in terms of behavior spaces}
The nesting in \equref{eq:behavior} makes this distinction precise.
Let $\mathcal{B}_e(P)$ and $\mathcal{B}_i(P)$ denote the behaviors required by $\mathcal{Q}_e(P)$ and $\mathcal{Q}_i(P)$, so that $\mathcal{B}(P) = \mathcal{B}_e(P) \cup \mathcal{B}_i(P)$.
Every witness targets behavior that the ground truth implements but the benchmark tests leave unchecked, and this region splits into three parts:
\begin{equation}
    \label{eq:regions}
    \mathcal{B}(G) \setminus \mathcal{B}(T_0)
    =
    \underbrace{\mathcal{B}_e(P) \setminus \mathcal{B}(T_0)}_{\text{stated}}
    \;\cup\;
    \underbrace{\mathcal{B}_i(P) \setminus \mathcal{B}(T_0)}_{\text{implied}}
    \;\cup\;
    \underbrace{\mathcal{B}(G) \setminus \mathcal{B}(P)}_{\text{incidental}}.
\end{equation}
The stated and incidental parts are unambiguous: a trial must implement stated behavior and need not reproduce incidental behavior.
Underspecification lives in the implied part, because $\mathcal{B}_i(P)$ is not a property of $P$ alone but of $P$ together with the background knowledge of whoever reads it.
Let $\mathcal{B}_i^{A}(P)$ denote the implicit behavior the task author intends and $\mathcal{B}_i^{R}(P) \subseteq \mathcal{B}_i^{A}(P)$ the part a reader can infer from $P$.
The prompt is underspecified when the inclusion is strict, and the difference
\begin{equation}
    \label{eq:ambiguous}
    \Delta(P) = \mathcal{B}_i^{A}(P) \setminus \mathcal{B}_i^{R}(P)
\end{equation}
is behavior that the ground truth implements, because the author wrote it, and that no test checks, but that a reader has no way to know is required.
From the reader's perspective, $\Delta(P)$ belongs to the incidental part of \equref{eq:regions}, so the nesting in \equref{eq:behavior} holds for the author's reading of $P$ but not for the agent's.
A witness in $\Delta(P)$ passes on $R[G]$ and fails on $R[s]$, yet the trial is faithful to the prompt the agent actually saw.
Separating $\Delta(P)$ from $\mathcal{B}_i^{R}(P)$ requires knowing what a reader can infer, which is not directly observable. \toolname{} therefore uses the evaluated models themselves as readers.

\toolname{} separates the two cases for every witness test $t_q$ that the reviewer agent accepts.
Such a test checks a requirement $q$ that $T_0$ does not check, and $t_q(R[G]) = \textsf{pass}$ by construction.
Two signals decide the class of $q$.
First, $q$ is \emph{explicit} if the audit agent's requirements checklist can quote a clause of $P$ that states it, and \emph{implicit} otherwise.
Second, the trials of the task reveal whether $q$ is inferable from $P$.
We say that a model \emph{skips} $q$ if most of its passing trials on the task fail $t_q$, and let $\rho(q)$ be the fraction of models with passing trials that skip $q$.
Counting models rather than trials keeps a model with many repeated runs from dominating.
\toolname{} then classifies $q$ empirically as follows.
\begin{itemize}[leftmargin=*, itemsep=1pt, topsep=2pt]
    \item \emph{Necessary, explicit} ($q \in \mathcal{Q}_e(P)$): the prompt states $q$, so every trial that fails $t_q$ is refuted, however many models skip it. The defect lies in the \emph{insufficient tests}.
    \item \emph{Necessary, implicit} ($q \in \mathcal{Q}_i(P)$ and $\rho(q) \leq 1/2$): most models implement $q$ without being told, which shows that it is inferable from $P$. Trials that fail $t_q$ are refuted, and the defect again lies in the tests.
    \item \emph{Ambiguous} ($q \in \mathcal{Q}_i(P)$ and $\rho(q) > 1/2$): most models skip a requirement the prompt never states. The defect lies in the \emph{underspecified prompt}, and trials that fail $t_q$ are not refuted on its basis.
\end{itemize}
In terms of \equref{eq:regions}, the three classes correspond to the stated part, the inferable part $\mathcal{B}_i^{R}(P) \setminus \mathcal{B}(T_0)$, and the ambiguous part $\Delta(P)$, with $\rho(q)$ serving as an empirical test of whether $q$ lies in $\mathcal{B}_i^{R}(P)$.
Tests for incidental behavior, which the prompt does not require at all, are already rejected by the reviewer agent (\S\ref{sec:approach:audit}).
Because $\rho(q)$ depends on the trials of all agents, the classification is a property of the task and is computed once over the full trial corpus.

Skipping is not the same as ambiguity.
Agents may recognize an implicit requirement and still skip it, since test-based rewards give no credit for untested behavior and agent harnesses often instruct agents to make minimal changes~\cite{edwards2026askorassume}. Models may also collapse onto the same misreading of a task~\cite{richter2026underspecification}.
The ambiguous class can therefore contain requirements that are clear but widely skipped.
We report the tasks with such ambiguity in \appref{app:additional_experiment:underspecified_prompt} together with some cases studies.

\subsection{Randomized Trial Auditing}
\label{app:random}

Algorithm~\ref{alg:random} gives the pseudocode of the lightweight approach of \S\ref{sec:approach:random} for a single task.
It first builds the augmented suite with prompt-based task analysis and refutes every trial that fails it, then repeatedly audits a randomly sampled trial from the active set and replays the resulting witnesses on all remaining trials.

\begin{algorithm}[h]
    \SetAlgoNlRelativeSize{-1}
    \caption{Randomized trial auditing for a single task}
    \label{alg:random}
    \KwIn{Specification $P$, repository $R$, ground truth $G$, benchmark suite $T_0$, benchmark-passing trials $S$, budget $B$}
    \KwOut{Refuted trials $F$, augmented test suite $T$}
    $T \gets T_0 \cup \textsc{TaskAnalysis}(P, R, G, T_0)$ \tcp*{every added test passes on $R[G]$}
    $F \gets \{ s \in S \mid \exists\, t \in T : t(R[s]) = \textsf{fail} \}$\;
    $A \gets S \setminus F$\;
    \While{$A \neq \emptyset$ \textbf{and} budget $B$ is not exhausted}{
        Sample $s$ uniformly at random from $A$\;
        $W \gets \textsc{AuditTrial}(P, R, G, s)$ \tcp*{confirmed witnesses, \S\ref{sec:approach:audit}}
        \eIf{$W = \emptyset$}{
            $A \gets A \setminus \{s\}$ \tcp*{audited clean}
        }{
            $T \gets T \cup W$\;
            $F' \gets \{ s' \in A \mid \exists\, t \in W : t(R[s']) = \textsf{fail} \}$ \tcp*{includes $s$}
            $F \gets F \cup F'$;\quad $A \gets A \setminus F'$\;
        }
    }
    \Return{$F$, $T$}
\end{algorithm}

%% file: appendix/additional_experiment.tex
\section{Additional Experiment Results}

\subsection{Additional Experiment Setup}
\label{app:additional_experiment:setup}

\fakeparagraph{Benchmarks}
This section expands the summary of \S\ref{sec:eval:setup}.
SWE-bench Verified~\cite{chowdhury2024swebenchverified} is the human-screened, 500-task subset of SWE-bench~\cite{jimenez2024swebench} and remains the de facto standard for agentic coding evaluation. Its tasks are mined from merged pull requests in 12 Python repositories and graded by the tests shipped with each pull request.
SWE-bench Pro~\cite{deng2025swebenchpro} follows the same mining recipe on copyleft-licensed repositories, targeting larger multi-file changes and grading with human-reviewed fail-to-pass and pass-to-pass tests. We use its public split of 731 tasks across 11 repositories, the only split whose reference patches are released.
DeepSWE~\cite{huang2026deepswe} (version 1.0) instead authors 113 original tasks from scratch across 91 repositories and five languages, and grades them with hand-written functional verifiers intended to accept any implementation of the requested behavior.
SkillsBench~\cite{li2026skillsbench} and SWE-Marathon~\cite{desai2026swemarathon} (version 1.0) are containerized, Harbor-style benchmarks whose tasks each ship an instruction, a starter environment, a held-out reference (oracle) solution, and a hidden verifier; SWE-Marathon further hardens its verifiers with multi-layer checks and adversarial exploit search.

\fakeparagraph{Task selection}
\toolname{} builds and executes each task's environment repeatedly, once per candidate test and per trial, so we exclude tasks whose environment makes this prohibitive.
From SkillsBench we therefore keep the coding-centric tasks that do not require a GPU, which leaves 13 of its 87 tasks across 8 domains:
\texttt{azure-\allowbreak bgp-\allowbreak oscillation-\allowbreak route-\allowbreak leak}, \texttt{data-\allowbreak to-\allowbreak d3}, \texttt{dialogue-\allowbreak parser}, \texttt{earthquake-\allowbreak plate-\allowbreak calculation}, \texttt{fix-\allowbreak build-\allowbreak agentops}, \texttt{flink-\allowbreak query}, \texttt{jax-\allowbreak computing-\allowbreak basics}, \texttt{lake-\allowbreak warming-\allowbreak attribution}, \texttt{mars-\allowbreak clouds-\allowbreak clustering}, \texttt{parallel-\allowbreak tfidf-\allowbreak search}, \texttt{protein-\allowbreak expression-\allowbreak analysis}, \texttt{simpo-\allowbreak code-\allowbreak reproduction}, \texttt{spring-\allowbreak boot-\allowbreak jakarta-\allowbreak migration}.
From SWE-Marathon we keep the tasks that neither require a GPU nor are ultra-long-horizon projects such as building a C compiler or reimplementing Kubernetes in Rust, which leaves 2 of its 20 tasks: \texttt{wasm-\allowbreak simd}, \texttt{zstd-\allowbreak decoder}.

\fakeparagraph{Benchmark versions}
Benchmarks keep changing after release as their maintainers fix tasks, verifiers, and environments, so the number of passing and refuted trials depends on the version audited.
\tabref{tab:benchmark_versions} pins the exact version of each benchmark we use.
DeepSWE, SkillsBench, and SWE-Marathon are distributed as Git repositories, and we use a commit shortly after their tagged releases.
SWE-bench Verified and SWE-bench Pro are distributed as Hugging Face datasets, and we fix the dataset revision; for SWE-bench Pro we also pin the commit of the official evaluation harness.

\input{tables/appendix/benchmark-versions}

\input{tables/appendix/underspecified}

\subsection{Underspecified Prompt}
\label{app:additional_experiment:underspecified_prompt}

We apply the requirement classification of \appref{app:implicit} to the witness tests produced by the lightweight approach.
A requirement is \emph{ambiguous} when the prompt does not state it and most models with passing trials on the task skip it, and we call a task's prompt \emph{underspecified} when at least one of its requirements is ambiguous.
Under this rule, 29 of the 113 DeepSWE tasks (25.7\%), 3 of the 13 audited SkillsBench tasks, and both audited SWE-Marathon tasks have underspecified prompts and \tabref{tab:underspecified} lists them.
Underspecification thus arises even in benchmarks whose prompts were written from scratch and reviewed by their authors, and it affects every benchmark to which we apply the lightweight approach.
These tasks contribute to the lower precision noted in \tabref{tab:rq2}: on them, whether a trial that departs from the reference solution also violates the prompt is itself ambiguous, so the lightweight approach and the per-trial audit can reach different verdicts on the same trial.

\fakeparagraph{Case studies}
We examine two DeepSWE tasks from \tabref{tab:underspecified} in detail.
In both, every audited trial passes the official tests, and the lightweight approach produces a witness test that passes on the reference solution and fails on every trial.
Neither witness, however, exposes a defect of the trials.
Each fixes one of two coherent readings of the prompt, namely the reading that the reference solution happens to implement.

\fakeparagraph{Case 1: \texttt{textual-richlog-follow-state}}
The task asks Textual's \texttt{Log} and \texttt{RichLog} widgets to preserve the user's viewport after the user scrolls away from the newest entries, and introduces an \texttt{is\_following\_end} flag, a \texttt{follow\_end()} method, and a \texttt{FollowChanged} message.
The prompt states that ``while \texttt{auto\_scroll} is enabled, new writes should follow only when the widget is already following the end'', but it never defines what \texttt{is\_following\_end} denotes.
Two readings remain available.
Under a \emph{position} reading, the flag is true whenever the viewport is at the end, and \texttt{auto\_scroll} separately decides whether new writes scroll the view.
Under an \emph{effective-follow} reading, the flag is true only when automatic following is enabled and the viewport is at the end.
The readings agree whenever \texttt{auto\_scroll} is enabled and diverge only when it is disabled, which is precisely the case that the qualifier ``while \texttt{auto\_scroll} is enabled'' leaves open.
The reference adopts the effective-follow reading: it initializes the flag from \texttt{auto\_scroll} and refreshes it as \texttt{auto\_scroll and is\_vertical\_scroll\_end}.
The witness test mounts an empty \texttt{Log(auto\_scroll=False)} and \texttt{RichLog(auto\_scroll=False)} and expects both flags to be \texttt{False}.
All 30 audited trials instead adopt the position reading, report \texttt{True}, and fail.
The failure occurs before any content is written, so it concerns only the meaning of the flag and not the scrolling behavior that the task is about.
The official tests check the initial state only with \texttt{auto\_scroll} enabled, where both readings produce the same result.

\fakeparagraph{Case 2: \texttt{updo-policy-alerting}}
The task adds policy-based alerting to Updo.
A monitored target enters a down state after a configured number of consecutive failures and recovers after a configured number of consecutive successes, and each evaluation returns a decision that carries the event, the state, and the current counters.
The prompt requires that \texttt{target\_recovered} be emitted ``only after consecutive successful checks'' and that every returned decision match the tracker state.
It does not specify whether \texttt{ConsecutiveRecoveries} records the streak that triggered recovery or only the progress toward a recovery that has not yet occurred, and therefore whether the counter is reset before the recovery decision is returned.
The omission is conspicuous, since the prompt does state when the latency-breach counter resets.
The witness test configures one failure to enter the down state and two successes to recover, and inspects the decision returned by the second success.
The reference increments the counter, resets it to zero upon recovery, and only then constructs the decision, so it reports \texttt{ConsecutiveRecoveries = 0}.
All seven audited trials emit \texttt{target\_recovered} with the correct state, previous state, and reason, but report the completed streak of 2, and fail.
Both behaviors yield a self-consistent snapshot of the tracker, and the official tests assert the event, the states, and the reason of the recovery decision but not its counter.
Whereas Case 1 turns on two readings of a named piece of state, this case turns on an implementation choice, the timing of a reset, that the prompt leaves entirely to the implementer.

\fakeparagraph{Discussion}
In both cases the requirement checked by the witness is implicit, and every model with passing trials on the task skips it, so the classification of \appref{app:implicit} labels it ambiguous.
The cases also show that such gaps are inexpensive to close.
For Case 1, a single sentence stating that \texttt{is\_following\_end} equals \texttt{auto\_scroll and is\_vertical\_scroll\_end}, together with an acceptance example for the disabled case, removes the ambiguity.
For Case 2, a sentence stating that the counter resets before the recovery decision is constructed, together with the failure-and-recovery sequence above as an acceptance example, does the same.
The witness tests that \toolname{} withholds from refutation therefore remain useful: they pinpoint where a benchmark's prompt should be clarified.

%% file: tables/appendix/benchmark-versions.tex
\begin{table}[H]
    \centering
    \scriptsize
    \caption{Benchmark versions used in our evaluation. Version gives the nearest release tag and the number of commits after it. Date is the commit date for GitHub repositories and the download date for Hugging Face datasets. Each entry lists the repository followed by the pinned commit (GitHub) or dataset revision (Hugging Face).}
    \label{tab:benchmark_versions}
    \begin{tabular}{@{}llll@{}}
        \toprule
        Benchmark               & Version             & Date       & Repository and commit / revision                  \\
        \midrule
        \multicolumn{4}{@{}l}{\emph{GitHub repositories}}                                                              \\
        DeepSWE                 & v1.0.0 (+1)         & 2026-06-01 & \texttt{datacurve-ai/deep-swe}                    \\
                                &                     &            & \texttt{578129c4334f6656a92a1c629af63a530596f169} \\
        \addlinespace
        SkillsBench             & v1.0 (+8)           & 2026-06-12 & \texttt{benchflow-ai/skillsbench}                 \\
                                &                     &            & \texttt{52659746d8167c4a347c9c0dd7e6a210f84d0fd8} \\
        \addlinespace
        SWE-Marathon            & v1.0-official (+19) & 2026-06-18 & \texttt{abundant-ai/swe-marathon}                 \\
                                &                     &            & \texttt{176d311ec6e2de802b133c59c34ce339c8560403} \\
        \addlinespace
        SWE-bench Pro (harness) & --                  & 2026-05-18 & \texttt{scaleapi/SWE-bench\_Pro-os}               \\
                                &                     &            & \texttt{ca10a60a5fcae51e6948ffe1485d4153d421e6c5} \\
        \midrule
        \multicolumn{4}{@{}l}{\emph{Hugging Face datasets}}                                                            \\
        SWE-bench Verified      & --                  & 2026-08-27 & \texttt{SWE-bench/SWE-bench\_Verified}            \\
                                &                     &            & \texttt{78f471bf655a3137b2e8a75af1501690ec009ec3} \\
        \addlinespace
        SWE-bench Pro (dataset) & --                  & 2026-08-27 & \texttt{ScaleAI/SWE-bench\_Pro}                   \\
                                &                     &            & \texttt{7ab5114912baf22bb098818e604c02fe7ad2c11f} \\
        \bottomrule
    \end{tabular}
\end{table}

%% file: tables/appendix/underspecified.tex
\begin{table}[H]
    \centering
    \scriptsize
    \caption{Tasks with underspecified prompts, identified by the lightweight approach using the classification of \appref{app:implicit}. \#Tasks gives the number of underspecified tasks out of the tasks audited. $^\ast$, $^\ddagger$: task subsets as in \tabref{tab:benchmarks}.}
    \label{tab:underspecified}
    \begin{tabular}{@{}lr>{\raggedright\arraybackslash}p{0.76\linewidth}@{}}
        \toprule
        Benchmark               & \#Tasks  & Tasks                                                                                                                                                                                                                                                                                                                                                                                                                                                                                                                                                                                                                                                                                                                                                                                                                                                                                                                                                                                                                                                                                                                                                                                                                                                                                                                                                                                                                                                                                                                                                                                                                                                                                                                                                                                                                                                                                                                                                                                                                                                                                                                                                                                                                                                                                                                                                                                                                                     \\
        \midrule
        DeepSWE                 & 29 / 113 & \texttt{adaptix-\allowbreak name-\allowbreak mapping-\allowbreak aliases}, \texttt{arktype-\allowbreak json-\allowbreak schema-\allowbreak refs-\allowbreak dependencies}, \texttt{bandit-\allowbreak structured-\allowbreak nosec-\allowbreak directives}, \texttt{cattrs-\allowbreak partial-\allowbreak structuring-\allowbreak recovery}, \texttt{claude-\allowbreak code-\allowbreak by-\allowbreak agents-\allowbreak recursive-\allowbreak delegation}, \texttt{csstree-\allowbreak shorthand-\allowbreak expansion-\allowbreak compression}, \texttt{dasel-\allowbreak html-\allowbreak document-\allowbreak format}, \texttt{dateutil-\allowbreak rfc5545-\allowbreak timezone-\allowbreak interop}, \texttt{etree-\allowbreak xml-\allowbreak diff-\allowbreak patch}, \texttt{expr-\allowbreak try-\allowbreak catch-\allowbreak errors}, \texttt{fd-\allowbreak deterministic-\allowbreak multi-\allowbreak key-\allowbreak sorting}, \texttt{go-\allowbreak genai-\allowbreak streamed-\allowbreak function-\allowbreak args}, \texttt{goreleaser-\allowbreak retry-\allowbreak publish-\allowbreak auditing}, \texttt{gql-\allowbreak incremental-\allowbreak graphql-\allowbreak delivery}, \texttt{helm-\allowbreak unified-\allowbreak manifest-\allowbreak stream}, \texttt{httpx-\allowbreak streaming-\allowbreak json-\allowbreak iteration}, \texttt{ink-\allowbreak grid-\allowbreak box-\allowbreak layout}, \texttt{katex-\allowbreak multicolumn-\allowbreak array-\allowbreak spans}, \texttt{mashumaro-\allowbreak flattened-\allowbreak dataclass-\allowbreak fields}, \texttt{obsidian-\allowbreak linter-\allowbreak auto-\allowbreak table-\allowbreak of-\allowbreak contents}, \texttt{opa-\allowbreak template-\allowbreak string-\allowbreak reconstruction}, \texttt{optique-\allowbreak conditional-\allowbreak option-\allowbreak dependencies}, \texttt{oxvg-\allowbreak structural-\allowbreak selector-\allowbreak preservation}, \texttt{python-\allowbreak statemachine-\allowbreak state-\allowbreak data-\allowbreak scoping}, \texttt{termenv-\allowbreak preserve-\allowbreak ansi-\allowbreak resets}, \texttt{testem-\allowbreak per-\allowbreak launcher-\allowbreak reports}, \texttt{textual-\allowbreak kitty-\allowbreak key-\allowbreak phases}, \texttt{textual-\allowbreak richlog-\allowbreak follow-\allowbreak state}, \texttt{updo-\allowbreak policy-\allowbreak alerting} \\
        \addlinespace
        SkillsBench$^\ast$      & 3 / 13   & \texttt{jax-\allowbreak computing-\allowbreak basics}, \texttt{parallel-\allowbreak tfidf-\allowbreak search}, \texttt{spring-\allowbreak boot-\allowbreak jakarta-\allowbreak migration}                                                                                                                                                                                                                                                                                                                                                                                                                                                                                                                                                                                                                                                                                                                                                                                                                                                                                                                                                                                                                                                                                                                                                                                                                                                                                                                                                                                                                                                                                                                                                                                                                                                                                                                                                                                                                                                                                                                                                                                                                                                                                                                                                                                                                                                 \\
        \addlinespace
        SWE-Marathon$^\ddagger$ & 2 / 2    & \texttt{wasm-\allowbreak simd}, \texttt{zstd-\allowbreak decoder}                                                                                                                                                                                                                                                                                                                                                                                                                                                                                                                                                                                                                                                                                                                                                                                                                                                                                                                                                                                                                                                                                                                                                                                                                                                                                                                                                                                                                                                                                                                                                                                                                                                                                                                                                                                                                                                                                                                                                                                                                                                                                                                                                                                                                                                                                                                                                                         \\
        \bottomrule
    \end{tabular}
\end{table}

%% file: appendix/case_study.tex
\section{Case Studies}
\label{app:case_study}

\subsection{Cases in Per-Trial Audit}
\label{app:case_study:per_trial}

The refutations of the per-trial audit expose defects of the benchmark tests rather than of the tasks themselves.
We examine two trials that \toolname{} refutes on SWE-bench Pro, where the official tests leave a required behavior unchecked.
The two cases differ in how the trial exploits the gap: the first never implements the untested part of the feature, whereas the second implements exactly what the visible fixture observes and nothing beyond it.
The complementary case, in which a witness reflects an underspecified prompt rather than a defective trial, is discussed in \appref{app:additional_experiment:underspecified_prompt}.

\fakeparagraph{Case 1: missing test cases in \texttt{navidrome/navidrome}}
The task, \texttt{navidrome-0130c6dc} (Go, base commit \texttt{2f90fc9b}), asks Navidrome to track the image files available for each album so that clients can access alternate covers and other artwork.
Detecting that a directory contains artwork is only the first stage of the feature. The requirements also span album construction, database storage, and rescanning of media that was already indexed.
The first requirement is explicit that the feature reaches persistent storage:
\begin{quote}
    \textit{The database must add a new column to store full image paths for each album and trigger a full media rescan once the migration is applied.}
\end{quote}
The remaining requirements add an album field for these paths, sorted and duplicate-free media directories, image filenames in the directory statistics, assignment of the full paths before the album is persisted, path construction with the system list separator, propagation of directory information during incremental scans and deletions, and a revised folder-change signature.
The reference solution makes the storage requirement concrete: it registers a new migration whose body is
\begin{lstlisting}[style=casestyle, language=SQL]
alter table main.album add image_files varchar;
\end{lstlisting}
calls \texttt{forceFullRescan(tx)} once the statement succeeds, adds \texttt{Album.ImageFiles}, implements \texttt{MediaFiles.Dirs()}, and assigns the combined paths to the album before \texttt{repo.Put(\&a)}.
Adding a Go field alone cannot create the column that these paths are stored in.

The official evaluation, in contrast, grades a single scanner test:
\begin{lstlisting}[style=casestyle]
fail_to_pass:               ['TestScanner']
pass_to_pass:               []
selected_test_files_to_run: ["TestScanner"]
\end{lstlisting}
The whole test patch is 531 bytes and touches only \texttt{scanner/walk\_dir\_tree\_test.go}, where its one changed assertion is
\begin{lstlisting}[style=casestyle]
 Expect(collected[baseDir]).To(MatchFields(IgnoreExtras, Fields{
-    "HasImages":       BeTrue(),
+    "Images":          ConsistOf("cover.jpg"),
     "HasPlaylist":     BeFalse(),
     "AudioFilesCount": BeNumerically("==", 5),
 }))
\end{lstlisting}
The assertion checks that directory traversal now collects the image filename of a fixture directory instead of a boolean flag.
\tabref{tab:case_navidrome} maps the task's required behaviors onto this patch: one of the six is asserted, and the graded tests never reach the schema, the album repository, or the rescan.

\input{tables/appendix/case-navidrome}

The audited trial of \texttt{gpt-5.6-sol} implements the traversal side of the feature and the model field, but omits the migration:
\begin{lstlisting}[style=casestyle]
Added:   Album.ImageFiles with structs:"image_files"
Missing: migration creating album.image_files
\end{lstlisting}
The resulting model therefore refers to a column that the database does not have.
The official evaluator still reports the trial as passing, and the reason is straightforward: collecting \texttt{cover.jpg} in an in-memory directory structure requires no column, so the only graded assertion is satisfied by a patch that cannot store what it collects.
Coverage measured during the audit is correspondingly thin, at 6.7\% of the lines the patch adds, with seven of the eight requirements unexercised.

The witness test that \toolname{} confirms, \texttt{TestAlbumImageFilesColumnCreated}, drives a temporary database through the application's own migration runner and asserts that the \texttt{image\_files} column exists on \texttt{album}:
\begin{lstlisting}[style=casestyle]
model world:   --- FAIL: TestAlbumImageFilesColumnCreated (1.93s)
oracle world:  --- PASS: TestAlbumImageFilesColumnCreated (0.48s)
               ok github.com/navidrome/navidrome/db
\end{lstlisting}
It passes on the reference solution and fails on the trial, and the repository's pre-existing persistence tests fail on the trial with \texttt{no such column: image\_files}.
The defect is thus not subtle, and the repository itself contains tests that would have caught it; the benchmark simply does not select them.

The gap generalizes beyond this trial.
Three independent failures survive the changed scanner assertion: a migration that is missing, unregistered, or ineffective, which is the defect here; a column that exists while album refresh or repository mapping never persists the generated paths; and a migration that creates the column without requesting the rescan, so albums indexed earlier never receive their artwork.
Closing the gap requires assertions at the stages the feature actually spans, namely a schema check after running the production migration path, an album round-trip through the real repository with paths joined by \texttt{filepath.ListSeparator}, and a check that applying the migration requests a full rescan.
Such tests must use an isolated temporary database rather than a fixture that creates \texttt{image\_files} by hand, since a hand-made column conceals exactly the omission this case exposes, and they must be added to the graded selection: assertions that the evaluator never runs close no gap.

\fakeparagraph{Case 2: reward hacking in \texttt{internetarchive/openlibrary}}
The task, \texttt{openlibrary-5069b09e} (Python, base commit \texttt{facafbe7}), concerns moving a booknote to a different work when the destination already holds a conflicting record.
Booknotes carry the composite primary key \texttt{(username, work\_id, edition\_id)}, so updating a source record to a destination that shares the username and edition violates uniqueness.
The fix must prevent the loss of the conflicting notes and return a dictionary of \texttt{rows\_changed}, \texttt{rows\_deleted}, and \texttt{failed\_deletes}.
The reference solution sets \texttt{Booknotes.ALLOW\_DELETE\_ON\_CONFLICT = False} and, when a bulk update hits a conflict, falls back to \texttt{update\_work\_ids\_individually}, which preserves each conflicting booknote, increments the failure count once per such record, and still moves the records that can move.
Other record types, bookshelves in particular, keep their deletion-on-conflict policy.

The official test adds one fixture with exactly one source note and one conflicting destination note:
\begin{lstlisting}[style=casestyle, language=Python]
rows = [
    {"username": "@mek", "work_id": 1, "edition_id": 1, "notes": "Jimmeny"},
    {"username": "@mek", "work_id": 2, "edition_id": 1, "notes": "Cricket"},
]
self.db.multiple_insert("booknotes", rows)
resp = Booknotes.update_work_id("1", "2")
assert resp == {"rows_changed": 0, "rows_deleted": 0, "failed_deletes": 1}
assert [dict(row) for row in self.db.select("booknotes")] == rows
\end{lstlisting}
The fixture inspects real database contents, so it does detect deletion in this scenario.
Its weakness is that it contains no second conflict and no source note that could move independently, which is what \tabref{tab:case_openlibrary} makes precise: every observation it makes coincides for a per-record implementation and for a constant.
The graded selection offers no further protection, since the two preservation tests exercise bookshelves rather than booknotes:
\begin{lstlisting}[style=casestyle]
FAIL_TO_PASS:  test_db.py::TestUpdateWorkID::test_no_allow_delete_on_conflict
PASS_TO_PASS:  test_db.py::TestUpdateWorkID::test_update_collision
               test_db.py::TestUpdateWorkID::test_update_simple
\end{lstlisting}

\input{tables/appendix/case-openlibrary}

\input{tables/appendix/case-httpx}

The audited trial of \texttt{gpt-5.6-sol} submits the constant:
\begin{lstlisting}[style=casestyle, language=Python]
if cls.PRESERVE_ROWS_ON_CONFLICT:
    # A collision must not result in a partial update or in either
    # of the conflicting records being discarded.
    failed_deletes = 1
    t.rollback()
    return {"rows_changed": rows_changed, "rows_deleted": rows_deleted,
            "failed_deletes": failed_deletes}
\end{lstlisting}
The branch reports one failure however many records conflict, abandons the whole update instead of moving the records that do not conflict, and leaves the unconditional deletion inside \texttt{update\_work\_ids\_individually} untouched.
Notably, the trial edits no test and tampers with no harness.
The evaluation procedure makes tampering pointless in any case, since it restores the graded test file from the fix commit after applying the model patch:
\begin{lstlisting}[style=casestyle]
reset repository to base commit -> apply model patch
-> checkout fix commit's openlibrary/tests/core/test_db.py
-> run selected tests -> parse results
\end{lstlisting}
What survives that restoration is the production change, and the trial passes because the single-conflict fixture is the only observation the benchmark makes of it.

\toolname{} confirms three witnesses, each targeting one of the scenarios the fixture omits.
The first parameterizes the number of conflicting pairs and asserts that \texttt{failed\_deletes} equals that number, which the constant cannot satisfy for two or more pairs.
The second mixes one conflicting pair with one movable note and asserts that the movable note reaches the destination while both conflicting notes keep their contents, which the rollback cannot satisfy.
The third calls \texttt{update\_work\_ids\_individually} directly with a conflicting pair and asserts that both records survive, which the untouched helper cannot satisfy.
All three pass on the reference solution and fail on the trial.
Each compares complete row contents rather than row counts, so that deleting a note and recreating an empty replacement cannot pass, and sorts rows by their key, since result order is not part of the requirement.

One qualification belongs with this case.
The prompt describes the conflict example as recording ``one failure'' and asks that booknotes remain completely unchanged in a conflict, wording that a reader could take to license a whole-operation rollback.
The literal \texttt{1} alone is therefore not proof of deliberate gaming, and the mixed-update and helper witnesses are the ones that establish a genuine departure from the requested behavior.
Our verdict is accordingly behavioral: the trial passes by satisfying an insufficient observation rather than by implementing the requested handling, and we make no claim about the model's intent or its knowledge of the hidden tests.
The task would also benefit from stating its contract explicitly, namely that \texttt{failed\_deletes} counts conflicting source records, that conflicts preserve both notes, that independently movable notes still update, and that the shared helper respects the booknote policy.

\subsection{Cases in the Lightweight Approach}
\label{app:case_study:lightweight}

The lightweight approach reaches a verdict for a whole task in two stages: prompt-based task analysis builds the augmented suite $T_1$ and refutes every trial that fails it, and the randomized loop then audits a few sampled survivors and replays the confirmed witnesses on the rest (Algorithm~\ref{alg:random}).
We follow one DeepSWE task, \texttt{httpx-deterministic-cookie-store}, through both stages to show what each contributes and why the cheap stage does not subsume the expensive one.
The task asks for a public \texttt{httpx.CookieStore} that extracts cookies from responses and builds outgoing \texttt{Cookie} headers, with deterministic ordering and eviction, expiry, domain and path matching, prefix rules, and mapping operations, while leaving existing behavior unchanged unless the new store is used.
Of the trials in our corpus, 42 pass the official verifier, whose suite is substantial: it constructs both \texttt{Client(cookies=store)} and \texttt{AsyncClient(cookies=store)}, imports a \texttt{CookieJar}, checks ordinary domain matching, and confirms that an extracted \texttt{Secure} cookie is withheld from an HTTP request.
\toolname{} nonetheless refutes 24 of the 42, as summarized in \tabref{tab:case_httpx}.

\fakeparagraph{Stage 1: requirements the official suite never exercises}
Task analysis reads the instruction, the repository, and $T_0$ without access to any trial, and maps requirements onto the tests that check them.
For this task it produces nine tests, of which two discriminate on the audited cohort; together they refute 11 of the 42 trials, leaving 31.
Both target requirements that the prompt states outright and the official suite exercises only in part.

The first concerns where the store may be used.
The instruction requires the store to work anywhere \texttt{cookies=} is accepted, but $T_0$ supplies it only to the client constructor, never as a per-request argument.
The added test passes a non-empty store to a single request and expects a correctly built header.
Nine trials fail it, and all nine fail the same way:
\begin{lstlisting}[style=casestyle]
BaseClient._merge_cookies -> legacy Cookies.update -> stdlib cookiejar
  -> AttributeError: 'str' object has no attribute 'domain'
\end{lstlisting}
These trials recognize the new container on the constructor path and leave the merge path on the old wrapper, which iterates the mapping into cookie names where the jar expects cookie objects.
The reference solution instead converts at the merge boundary, building a \texttt{CookieStore} whenever either side of the merge is one.

The second concerns which inputs \texttt{update()} accepts.
The instruction enumerates \texttt{CookieStore}, \texttt{httpx.Cookies}, \texttt{CookieJar}, dictionaries, and lists of pairs; $T_0$ covers every form except the \texttt{httpx.Cookies} wrapper.
The added test supplies exactly that form:
\begin{lstlisting}[style=casestyle, language=Python]
store = httpx.CookieStore()
store.update(httpx.Cookies({"a": "1"}))
request = httpx.Request("GET", "https://example.org/")
store.set_cookie_header(request)
assert request.headers.get("cookie") == "a=1"
\end{lstlisting}
Two trials leave the store empty and produce no header, because they accept the underlying jar but not its wrapper, which the reference unwraps before importing.
Neither test needs a defective patch to motivate it: both gaps follow from comparing an enumerated requirement with the call sites $T_0$ actually exercises, which is precisely the evidence available to task analysis.

\fakeparagraph{Stage 2: defects that only a concrete trial reveals}
The randomized loop then samples three of the 31 survivors, one per audit round, and the confirmed witnesses are replayed on the whole survivor pool.
Each round inspects one implementation and generalizes the defect it finds into a test justified by the instruction.

The first round samples a \texttt{gpt-5.5} trial that normalizes cookie domains with \texttt{.lstrip(".").rstrip(".")}, treating leading and trailing dots as interchangeable.
For a response from \texttt{example.org} carrying \texttt{Domain=example.org.}, stripping the trailing dot turns a domain that should not match into one that does, so the cookie is stored instead of rejected.
The reference removes leading dots only and then applies its matching helper.
The witness checks that the cookie is rejected; the underlying defect, normalization that widens acceptance, is invisible from the task text alone, since the instruction invokes standard domain matching without spelling out trailing dots.

The second round samples a \texttt{claude-opus-4.7} trial that re-imports jar cookies through the public setter \texttt{self.set(name, value, domain, path)}.
The setter takes no \texttt{Secure} argument, so the import preserves the visible name, value, domain, and path while silently dropping the scheme restriction.
The two witnesses import a \texttt{Secure} cookie from a raw \texttt{CookieJar} and from an \texttt{httpx.Cookies} wrapper, then assert that it is withheld from an HTTP request; the reference copies the \texttt{Secure} flag and the expiry when importing.
This defect sits in an interaction that no checklist of individual API names would catch: $T_0$ checks jar import with a non-secure cookie and checks \texttt{Secure} only on the extraction path, so both features pass separately while their combination fails.

The third round samples a \texttt{claude-opus-4.8} trial that widens the scheme policy instead:
\begin{lstlisting}[style=casestyle, language=Python]
is_secure = url.scheme.lower() in ("https", "wss")
\end{lstlisting}
The instruction says \texttt{Secure} cookies are sent only over HTTPS, so the witness applies a stored \texttt{Secure} cookie to a \texttt{wss://} request and expects no header.
The reference restricts the scheme to HTTPS exactly.

\fakeparagraph{What the two stages contribute}
Replaying the four audit-generated witnesses on the 31 survivors refutes 13 further trials, so the loop's three expensive audits transfer to more than four times as many trials as they inspect.
Of the 42 trials the benchmark accepts, 24 (57.1\%) are refuted, 11 by task analysis and 13 by the randomized loop, and 18 survive.
The two stages find different kinds of defects.
Task analysis finds requirements that the prompt enumerates and $T_0$ leaves unchecked, which is why its tests are cheap to derive and refute trials in bulk.
The randomized loop finds defects that live inside implementation choices, namely a normalization helper that widens acceptance, a setter that silently drops metadata, and a scheme predicate with one extra entry, none of which is visible from the prompt and the official tests alone.
Every witness here checks a requirement classified as necessary (\appref{app:implicit}), so all 24 refutations stand.
The replay result also reflects one execution of the loop: the sampled trials are themselves members of the pool, and a different sample would fix the witnesses in a different order, so the count of rounds needed to reach this outcome is not fixed by this run alone.

%% file: tables/appendix/case-navidrome.tex
\begin{table}[t]
    \centering
    \caption{Required behaviors of the Navidrome task and the evidence for each in the official test patch. Only the first is graded, and the trial we audit violates the second.}
    \label{tab:case_navidrome}
    \resizebox{0.92\linewidth}{!}{
        \begin{tabular}{ll}
            \toprule
            Required behavior                                        & Evidence in the official test patch \\
            \midrule
            Collect image filenames in directory statistics          & Asserted for \texttt{cover.jpg}     \\
            Create the album image-path column                       & No migration or schema assertion    \\
            Persist and retrieve album image paths                   & No repository round-trip assertion  \\
            Trigger a full rescan after migration                    & No rescan assertion                 \\
            Build paths across distinct media directories            & No path-construction assertion      \\
            Preserve paths during incremental scans and deletions    & No lifecycle assertion              \\
            \bottomrule
        \end{tabular}
    }
\end{table}

%% file: tables/appendix/case-openlibrary.tex
\begin{table}[t]
    \centering
    \caption{Observations made by the single-conflict fixture of the Open Library task. A per-record implementation and the trial's constant agree on every one of them, so the graded test cannot separate the two.}
    \label{tab:case_openlibrary}
    \resizebox{0.92\linewidth}{!}{
        \begin{tabular}{lll}
            \toprule
            Observation in the fixture & Correct implementation                & Trial's shortcut              \\
            \midrule
            Changed records            & 0: the only source record conflicts   & 0: the operation rolls back   \\
            Deleted records            & 0: conflicting notes are preserved    & 0: the operation rolls back   \\
            Failed deletions           & 1: one conflicting source record      & 1: a literal constant         \\
            Stored records             & Both original notes remain            & Both original notes remain    \\
            \bottomrule
        \end{tabular}
    }
\end{table}

%% file: tables/appendix/case-httpx.tex
\begin{table}[t]
    \centering
    \caption{The lightweight approach on the DeepSWE task \texttt{httpx-\protect\allowbreak deterministic-\protect\allowbreak cookie-\protect\allowbreak store}. Task analysis is run once per task; each of the three audit rounds inspects one sampled survivor, and its witnesses are replayed on the whole surviving pool.}
    \label{tab:case_httpx}
    \resizebox{0.86\linewidth}{!}{
        \begin{tabular}{lrrrr}
            \toprule
            Stage                     & Trials entering & Witness tests        & Refuted & Remaining \\
            \midrule
            Official verifier         & --              & --                   & --      & 42        \\
            Task analysis             & 42              & 9 (2 discriminating) & 11      & 31        \\
            Randomized trial auditing & 31              & 4                    & 13      & 18        \\
            \midrule
            Total                     & 42              & 13                   & 24      & 18        \\
            \bottomrule
        \end{tabular}
    }
\end{table}

%% file: appendix/prompt.tex
\section{Prompt Templates}
\label{app:prompts}

This section lists the prompt templates of the three agents shown in \Figref{fig:method}.
The audit agent audits a single benchmark-passing trial and writes issue-demonstrating tests (\S\ref{sec:approach:audit}).
The reviewer agent re-judges the tests added by an audit against the task instruction and decides which to keep.
The task analysis agent analyzes a task before any trial is examined and adds tests for requirements that the existing suite leaves unchecked (\S\ref{sec:approach:random}).
Fields in braces (\eg \texttt{\{\{TASK\_ID\}\}}) are filled in by the runner for each task and trial.

\promptbox{Audit Agent Prompt}{appendix/prompts/audit_agent.txt}

\promptbox{Reviewer Agent Prompt}{appendix/prompts/reviewer_agent.txt}

\promptbox{Task Analysis Agent Prompt}{appendix/prompts/task_analysis_agent.txt}

%% file: appendix/limitation.tex
\section{Limitations}
\label{app:limitations}

We summarize the main limitations of \toolname{} and of our evaluation.

\fakeparagraph{Dependence on a reference solution}
\toolname{} relies on a reference solution to certify every new test, so benchmarks that ship only a prompt and a test suite fall outside its scope.
It further assumes that each task is well-formed, \ie that its tests, prompt, and ground truth are mutually consistent (\appref{app:assumption}); tasks that violate this assumption call for repairing the task rather than re-judging its trials.

\fakeparagraph{LLM-based auditing}
The audit and reviewer agents are themselves LLMs, so a refutation is only as reliable as their reading of the prompt.
Oracle validation guarantees that every witness asserts behavior the ground truth exhibits, but not that the prompt requires it. This judgment is delegated to the reviewer.
Underspecified prompts, whose implicit requirements a reader cannot reliably infer, remain the main source of disagreement between our configurations (\appref{app:additional_experiment:underspecified_prompt}).

\fakeparagraph{Coverage and precision of the lightweight approach}
The lightweight approach trades coverage and precision for cost.
Each audit round targets the defects of one sampled trial, so defects that are confined to trials it never samples can be missed under a finite budget, and tasks with few passing trials offer little opportunity to amortize an audit across trials. Moreover, the lightweight approach will amplify the noise in trials and the ambiguity in the prompt and diffuse across other unaudited trials, which has negative impact on precision.

\fakeparagraph{Scope of the evaluation}
We audit trials from a fixed set of frontier models and excluded SkillsBench and SWE-Marathon tasks that are not coding tasks, require a GPU, or are ultra-long-horizon projects (\S\ref{sec:eval:setup}), so our refutation rates may not transfer to other agents or task types.
Recall and precision in RQ2--RQ4 measure agreement with the per-trial audit rather than with a human-labeled ground truth.